\documentclass[a4paper,aps,prl,preprint,superscriptaddress,showpacs]{revtex4}

\usepackage{graphicx}
\usepackage{color}
\usepackage{amsmath}
\usepackage{enumerate}

\begin{document}

\title{The determination of the Dzyaloshinskii-Moriya interaction in exchange biased Au/Co/NiO system}

\author{P.~Ku\'{s}wik}
\email{kuswik@ifmpan.poznan.pl}
\affiliation{Institute of Molecular Physics, Polish Academy
of Sciences, Smoluchowskiego 17, 60-179 Pozna\'{n},~Poland}
\author{M.~Matczak}
\affiliation{Institute of Molecular Physics, Polish Academy
of Sciences, Smoluchowskiego 17, 60-179 Pozna\'{n},~Poland}
\author{M.~Kowacz}
\affiliation{Faculty of Technical Physics, Pozna\'{n} University of Technology, Piotrowo 3, 60-965 Pozna\'{n},~Poland}
\author{F. Lisiecki}
\affiliation{Institute of Molecular Physics, Polish Academy
of Sciences, Smoluchowskiego 17, 60-179 Pozna\'{n},~Poland}

\author{F.~Stobiecki}
\affiliation{Institute of Molecular Physics, Polish Academy
of Sciences, Smoluchowskiego 17, 60-179 Pozna\'{n},~Poland}
\affiliation{NanoBioMedical Centre, Adam Mickiewicz University, Umultowska 85, 61-614 Pozna\'{n},~Poland}

\date{\today}

\begin{abstract}
The determination of the interfacial Dzyaloshinskii-Moriya interaction for perpendicularly magnetized thin layered films is not trivial and therefore needs some  experimental efforts to determine its sign and magnitude, especially in the exchange biased system. Here, we developed a method proposed by D.-S. Han et al. [Nano Lett. 16, (2016) 4438], which opens a way to investigate this interaction using a conventional PMOKE magnetometer, also in exchange biased systems. Using our approach we demonstrated that the Au/Co/NiO layered system has a strong negative Dzyaloshinskii-Moriya interaction, which is independent of the direction of perpendicular interlayer exchange bias coupling.
 
\end{abstract}


\maketitle
\section {Introduction}
Nowadays, the chiral spin configuration  is widely investigated in perpendicularly magnetized thin layered films with the interfacial Dzyaloshinskii-Moriya interaction (DMI)~\cite{Sampaio2013,Boulle2016,Moreau-Luchaire2016} due to possible applications in spintronic devices based on new concepts (e.g.,  bubblecade ~\cite{Moon2015} or racetrack memory~\cite{Parkin190,Emori2013, Tomasello2014}). In such devices material parameters, especially the strength and sign of DMI, play an essential role. To determine DMI, the most common method is to measure the magnetization dynamics in the GHz regime using Brillouin light scattering  (BLS)~\cite{Cho2015,Boulle2016,doi:10.1063/1.4945685} or Vector Network Analyzer – Ferromagnetic Resonance (VNA-FMR)~\cite{Lee2016}. Both methods require  sophisticated measuring devices and are challenging to distinguish DMI in very thin ferromagnetic (FM) films with a perpendicular magnetic anisotropy (PMA). Moreover, using these methods the different interfacial anisotropies located on the top and the bottom interfaces of FM layer significantly hinder analysis of the results. The interpretation could be especially difficult, when large additional contribution to the surface anisotropy is present only at one interface (e.g., for exchange biased  AFM/FM system~\cite{doi:10.1063/1.4952706,PhysRevB.91.134413}).
 
Another approach to determine the DMI is an observation of domain walls (DWs) and their asymmetric propagation under the in-plane magnetic field ($H_{\rm{x}}$). In this case the domain wall velocity versus $H_{\rm{x}}$ is analyzed based on the creep law, where  $v$($H_{\rm{x}}$) dependance is shifted with a minimum occurring for $H_{\rm{x}}$ equals to intrinsic DMI field \cite{Hrabec2014}. However, for some systems (eg., Pt/Co/GdO$_{\rm{x}}$),  this $v$($H_{\rm{x}}$) cannot be explained by the variation of the DW energy with $H_{\rm{x}}$ in the creep regime. Therefore, to obtain the strength and the sign of the DMI, the flow regime of propagation should be analyzed~\cite{0953-8984-27-32-326002}. Moreover,  $v$($H_{\rm{x}}$) dependance often shows shift and an asymmetric shape, analysis of which requires consideration of the other factors (e.g., chiral damping or variation of anisotropy~\cite{Jue2015,Lau2016}) making the analysis of DMI much more complicated. 

Very recently D.-S. Han et al.~\cite{Han2016} have developed a promising method, which overcomes mentioned difficulties. This approach is based on  the analysis of magnetization reversal process in triangular-shaped microstructures in a sweeping perpendicular magnetic field ($H_{\rm{z}}$) with an additional constant in-plane magnetic field ($H_{\rm{x}}$), where asymmetric hysteresis loops are observed. The hysteresis loops were obtained from domain evolution recorded with  Polar Magnetooptical Kerr (PMOKE) microscope. Analysis of the shift of the magnetic hysteresis loops, based on a droplet model ~\cite{Han2016,Pizzini2014} for domain nucleation, enables the determination of the strength and sign of interfacial DMI.  However, this approach requires the use of sophisticated PMOKE microscope with high spatial resolution of magnetic domain structure in a single triangle. An additional difficulty in these measurements is a small magnetooptical signal. To overcome this disadvantage the results have to be collected from a single pattern over 10 times and then needs to be fitted to obtain switching fields.

In this paper we propose a modified method presented by D.-S. Han et al.~\cite{Han2016}. In our approach, the array of 156 triangular-shaped microstructures located on the 100$\times$100~$\mu$m$^2$ surface area are investigated using the PMOKE magnetometer. The hysteresis loops were collected in the $H_{\rm{z}}$ sweeping field with a constant $H_{\rm{x}}$. This enables us to determine the sign and the magnitude of DMI for exchange biased system, for which obtaining the DMI by other techniques is difficult. Here, we measured  exchange biased Au/Co/NiO  layer system, where the N\'eel domain walls with clockwise chirality was found independently on direction of interlayer exchange bias coupling (IEBC)~\cite{Kus2017b}. Using our approach we were able to confirm that this domain wall configuration is caused by a strong negative DMI, which sign and magnitude is independent of the IEBC direction.

\section {Experimental}

To fabricate the triangular microstructure we used the EBL technique based on the positive resist spin coated onto naturally oxidized Si substrate. In the lift-off method we deposited the Ti-4nm/Au-60nm/Co-0.8nm/NiO-10nm/Au-2nm structure using magnetron sputtering (Co, Au, and Ti layers) and pulsed laser deposition (NiO layer) in the external magnetic field of $H$$_{\rm{dep}}$=1.1 kOe, oriented perpendicularly to the sample plane. Deposition in external magnetic field enables us to establish the perpendicular IEBC between Co and NiO layers~\cite{doi:10.1063/1.4952706}. The detailed description of the deposition procedure is given in our previous papers~\cite{doi:10.1063/1.4952706, Kus2017b}. The quality of the structure was controlled by using a scanning electron microscope (SEM). To investigate the role of the IEBC direction on the DMI we additionally performed field cooling (FC) procedure from 353~K cooled down to RT in two different external magnetic fields $H$$_{\rm{FC^{-}}}$=-2 kOe and $H$$_{\rm{FC^{+}}}$=+5 kOe applied perpendicularly to the sample plane. PMOKE hysteresis loops were measured using a magnetometer with a laser diode (wavelength 655 nm and the beam diameter $<$0.3 mm).

\section {Results and disscusion}

\begin{figure}[hp]
\centering \vspace*{-0.3cm}\includegraphics[scale=0.6]
{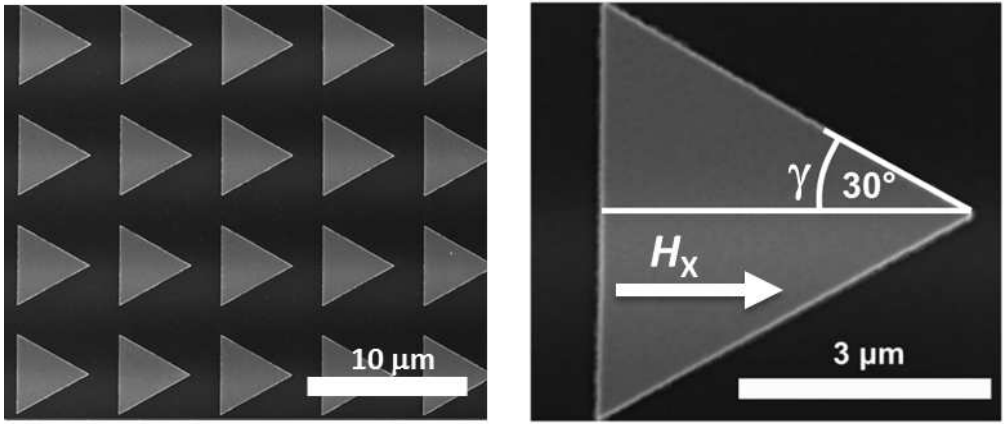} \vspace*{-0.2cm} \caption{The SEM images of a) part of the array of triangles and b) the single triangle microstructure demonstrating good quality of the edges after lift-off process.}\label{SEM}
\end{figure}

An important aspect of this investigation was to fabricate the microstructures with smooth edges, because a difference between nucleation barrier at each side of the triangle was used to obtain the DMI and this nucleation process cannot be affected by topological defects. Therefore, the quality of the triangles edges were verified by the SEM (Fig.~\ref{SEM}). Using standard lift-off lithographic technique we were able to fabricate the array of 156 triangle elements (Fig.~\ref{SEM}a) characterized by high shape repeatability. In Fig.~\ref{SEM}b we have shown that obtained microstructures have smooth edges similar to that presented in Ref.~\cite{Han2016}. Note, that in this work we investigated equilateral triangles, in which triangle edges and constant in-plane magnetic field, oriented perpendicularly to the base of the triangle, ($H_{\rm{x}}$) form an angle ($\gamma$)  (Fig.~\ref{SEM}b). 
\begin{figure}[tbp]
\centering \vspace*{-0.3cm}\includegraphics[scale=0.6]
{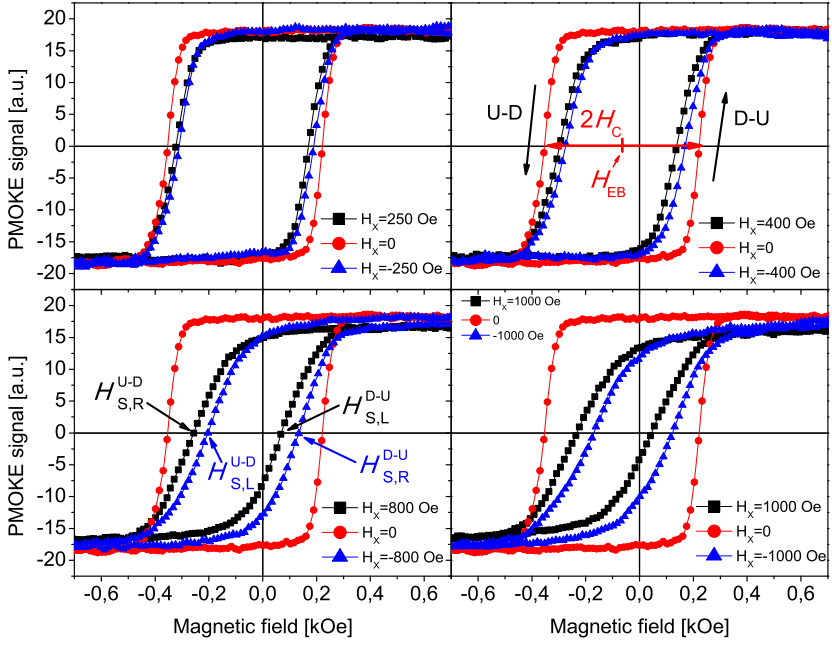} \vspace*{-0.2cm} \caption{PMOKE hysteresis loops under constant $H_{\rm{x}}$ field measured for the array of triangles of Ti-4nm/Au-60nm/Co-0.8 nm/NiO-10nm/Au-2nm in as-deposited state.}\label{Histeresis}
\end{figure}

To determine the sign and magnitude of DMI for exchange biased Ti-4nm/Au-60nm/Co-0.8nm/NiO-10nm/Au-2nm layered system we measured the hysteresis loops (Fig.~\ref{Histeresis}) for the array of triangles in sweeping $H_{\rm{z}}$ field with constant $H_{\rm{x}}$ (Fig.~\ref{SEM}b). In our measurements the magnetooptical signal-to-noise ratio is very high and thus magnetization switching fields from down-to-up ($H_{\rm{S}}^{\rm{D-U}}$) and up-to-down ($H_{\rm{S}}^{\rm{U-D}}$) can be easily read without any additional averaging. Note that $H_{\rm{S}}^{\rm{D-U}}$ and $H_{\rm{S}}^{\rm{U-D}}$ can be distinguished, depending on which sides of the triangles the domination of nucleation appears ($H_{\rm S,L}$ and $H_{\rm S,R}$ denotes left or right-side of triangles, respectively). For the hysteresis loop measurements, the symmetric feature of tilted magnetization on the edges does not break in the case of $H_{\rm{x}}$=0, therefore the shift along $H_{\rm{z}}$ field ($H_{\rm shi}$=$\frac{H_{\rm{S}}^{\rm{D-U}}+H_{\rm{S}}^{\rm{U-D}}}{2}$) reveals the presence of IEBC between the Co and NiO layers. However, when $H_{\rm{x}}$$\neq$0, additional shift appears, since the DMI caused tilting of the local edge magnetization in respect to the magnetization inside the triangles. As a consequence, the different Zeeman energy between left and right edges leads to a reduction in energy barrier for the magnetization reversal in $H_{\rm{z}}$. This tilt direction is related to the sign of DMI and thus the hysteresis shift is used to determine the direction of Dzyaloshinskii-Moriya vector~\cite{Han2016}. Because in our system additional shift is induced by IEBC, we will take it into account in the analysis of DMI. Note that in our sample the hysteresis loop is slightly sloped, indicating that small differences in the nucleation fields between each triangle may occur.

 \begin{figure}[tbp]
\centering \vspace*{-0.3cm}\includegraphics[scale=0.6]
{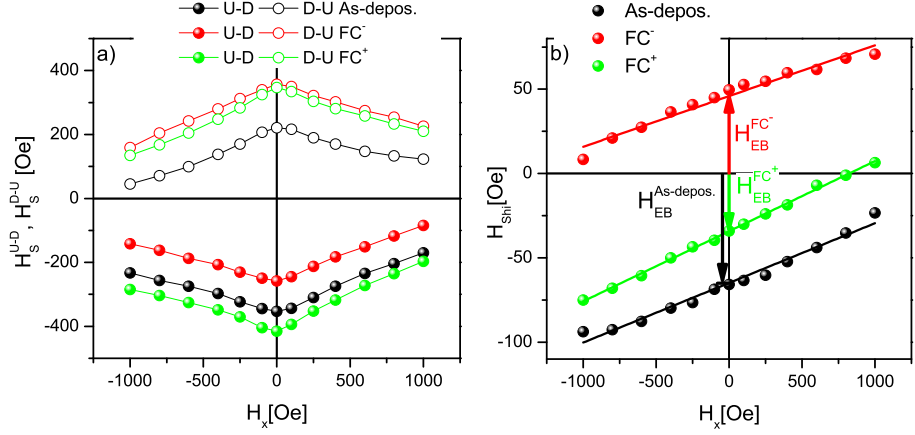} \vspace*{-0.2cm} \caption{a) Magnetization switching field for D-U and U-D process and b) dependance of $H_{\rm shi}$ versus $H_{\rm x}$ for Ti-4nm/Au-60nm/Co-0.8 nm/NiO-10nm/Au-2nm in as-deposited, FC$^+$ and FC$^-$ state. On b) $H_{\rm{EB}}$ is marked for each state.}\label{HEB}
\end{figure}

Performing an additional FC$^+$ and FC$^-$ process we were able to change the direction of the IEBC~\cite{Kus2017b}. This allowed to check the influence of IEBC on hysteresis loop shift induced by DMI. Previously we have demonstrated ~\cite{Kus2017b} that the FC processes performed in an opposite magnetic field direction does not change the chirality of the N\'eel domain walls, therefore we expect that the shift direction will be the same after different FC process. Indeed, the hysteresis loop shift direction is the same in all three cases (as-deposited, FC$^+$ and FC$^-$) (Fig.~\ref{HEB}b), however, the switching fields $H_{\rm{S}}^{\rm{D-U}}$($H_{\rm x}$) and $H_{\rm{S}}^{\rm{U-D}}$($H_{\rm x}$) are different (Fig.~\ref{HEB}a). Therefore, we have to clarify whether the IEBC influences $H_{\rm{S}}^{\rm{D-U}}$ and $H_{\rm{S}}^{\rm{U-D}}$, which are affected by DMI. To do that, we analyze the $H_{\rm shi}$($H_{\rm x}$) dependence, which shows a linear behavior (Fig.~\ref{HEB}b). These lines are almost parallel to each other independently of the sign of the $H_{\rm EB}$, where the intercept is equal to the $H_{\rm EB}$=$H_{\rm shi}$($H_{\rm x}$=0). Based on these results we assumed that $H_{\rm EB}$ does not change during magnetization reversal versus $H_{\rm x}$. Analyzing these data we need to be sure that the $H_{\rm EB}$ is not changing during the recording of hysteresis loop in $H_{\rm z}$, therefore prior to the measurements we performed 20 cycles of magnetization reversal to minimize the training effect~\cite{Radu2008}. Taking into account this assumption and the description of the nucleation process for triangle-shaped microstructure~\cite{Han2016}, we can write the normalized switching fields for U-D and D-U magnetization reversal process including $H_{\rm EB}$: 
 
\begin{equation}
\label{eq:1}
\frac{H_{\rm S,L}-H_{\rm EB}}{H_{\rm C}}=\frac{\sigma^2 (H_{\rm x})}{\sigma^2 (0) \sqrt{1-(\frac{H_{\rm x}}{H_{\rm K}})^2}},~~\frac{H_{\rm S,R}-H_{\rm EB}}{H_{\rm C}}=\frac{\sigma^2 [H_{\rm x} \rm sin(\gamma)]}{\sigma^2 (0) \sqrt{1-(\frac{H_{\rm x}}{H_{\rm K}})^2}},
\end{equation}

where $H_{\rm C}$ is a coercive field for $H_{\rm x}$=0 and $H_{\rm K}$ is an in-plane saturation field. Here, the $\sigma(H_{\rm x})$ is a domain wall energy modified by the DMI expressed as $\sigma(H_{\rm x})=\sigma_{\rm 0} [\sqrt{1-({\frac{H_{\rm x}}{H_{\rm K}}})^2} - (\frac{H_{\rm x}}{H_{\rm K}}+\frac{2D}{\sigma_{\rm 0}})\rm arccos(\frac{H_{\rm x}}{H_{\rm K}})]$, where $\sigma_{\rm 0}=4\sqrt{\rm A K_{\rm eff}}$ describes  the domain wall energy without DMI in $H_{\rm x}$=0 ($A$-exchange stiffness, $K_{\rm eff}$- effective magnetic anisotropy).

\begin{figure}[hbp]
\centering \vspace*{-0.3cm}\includegraphics[scale=0.4]
{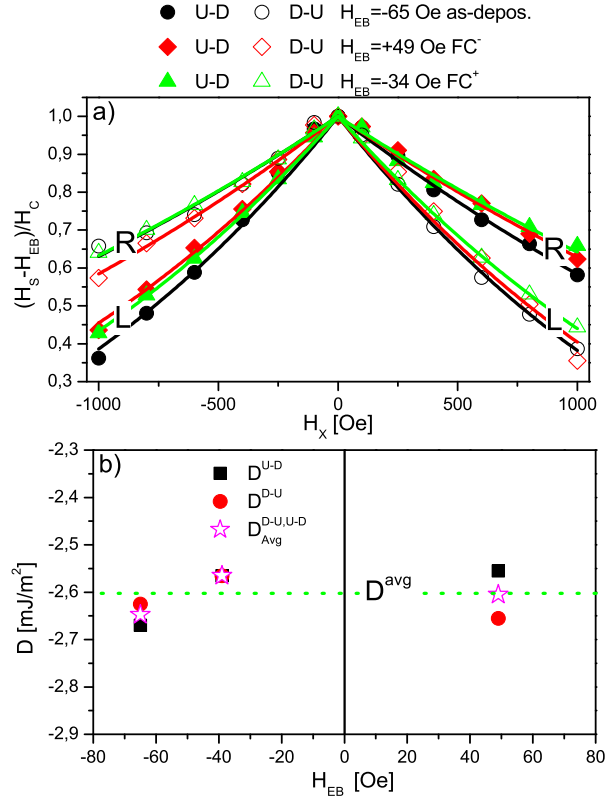} \vspace*{-0.2cm} \caption{a) Normalized magnetization switching fields for D-U and U-D process in the $H_{\rm x}$ and b) determined values of DMI for Ti-4nm/Au-60nm/Co-0.8nm/NiO-10nm/Au-2nm in as-deposited, FC$^+$ and FC$^-$ state, where R and L denotes the nucleation induced on the right and left triangles edge side, respectively. Lines in a) indicate fit to Eq.\ref{eq:1}.}\label{D}
\end{figure}  

To fit the Eq.~\ref{eq:1} to our experimental data (Fig.~\ref{D}) we assumed that A=10 pJ/m, which is the value used for the Co layer in similar systems~\cite{Han2016,Moreau-Luchaire2016}. Other material parameters were determined from our previous magnetic measurements for Au/Co/NiO/Au layer system~\cite{doi:10.1063/1.4952706}: $H_{\rm K}$= 12.2 kOe and $K_{\rm eff}$=0.69 MJ/m$^3$. 

Focusing on the value of the DMI in the Au/Co/NiO layered system we performed a fitting procedure for the four branches of normalized switching fields for D-U and U-D magnetization reversal process in the $+H_{\rm x}$ and $-H_{\rm x}$ (lines in Fig.~\ref{D}a). From theses fits the value of the $D^{\rm{U-D}}$ and $D^{\rm{D-U}}$ was extracted for as-deposited and FC samples (Fig.~\ref{D}b).  It is clear that $D^{\rm{U-D}}$ and $D^{\rm{D-U}}$  for a single stage (as-deposited, FC$^+$ and FC$^-$) can be fitted with almost the same value, supporting interpretation derived from magnetic domain evolution given in Ref.~\cite{Kus2017b} that DMI ($D^{\rm{avg}}$=-2,6 mJ/m$^2$) is independent of IEBC direction, and therefore can be tuned independently of the IEBC.

Moreover, using this method we also verified the correlation beetwen the N\'eel domain walls with clockwise chirality with the effective sign of DMI in the Au/Co/NiO layer system. For this chirality we expect a negative DMI~\cite{Yang2015}, therefore the nucleation process for U-D (D-U)  should be induced on the right (left) edge side, where the Zeeman energy barrier is a lowered when a direction of the local edge magnetization forms lower angle with $H_{\rm x}$. As a result, the hysteresis loop should be shifted along $-H_{\rm{z}}$ field for $+H_{\rm x}$ and in opposite direction for $-H_{\rm x}$. Indeed,  we have found this behavior for hysteresis loops measured with various $H_{\rm x}$, what unambiguously confirms that in the exchange biased Au/Co/NiO layered films the DMI is negative.

\section {Summary}
In this work we showed a modified method proposed by D.-S. Han et al. [Nano Lett. 16, (2016) 4438], which opens a way to investigate the interfacial Dzyaloshinskii-Moriya interaction using a conventional PMOKE magnetometer. This approach allows to determine both the sign and the magnitude of Dzyaloshinskii-Moriya interaction in exchange biased thin films. Using this method we demonstrated that the Au/Co/NiO layered system has strong negative DMI, which is independent of IEBC direction.

\section {Acknowledgments}
The authors thank Professor J. Korecki for valuable discussions.
This work was  supported by the National Science Center Poland partially under the SONATA-BIS (UMO-2015/18/E/ST3/00557) and the MAESTRO fundings (UMO-2011/02/A/ST3/00150).

\end{document}